\documentclass[]{spie}  

\usepackage{amsmath,amsfonts,amssymb}
\usepackage{graphicx}
\usepackage[colorlinks=true, allcolors=blue]{hyperref}
\usepackage[perpage]{footmisc} 
\usepackage{nicefrac}
\usepackage{subcaption}
\usepackage{csquotes}

\newcommand{\reffig}[1]{Figure \ref{#1}}
\newcommand{\refsec}[1]{Section \ref{#1}}
\newcommand{\refnum}[1]{Ref.~\citenum{#1}}
\newcommand{\nlhref}[1]{\href{#1}{\nolinkurl{#1}}} 

\title{The Nancy Grace Roman Space Telescope Coronagraph Community Participation Program}

\author[a,b]{Dmitry Savransky}
\affil[a]{Sibley School of Mechanical and Aerospace Engineering, Cornell University, Ithaca, NY, 14853, USA}
\affil[b]{Carl Sagan Institute, Cornell University, Ithaca, NY, 14853, USA}

\author[c]{Vanessa P. Bailey}
\author[d]{Schuyler G. Wolff}
\author[e]{Maxwell A. Millar-Blanchaer}
\author[f]{Jason Wang}
\author[g]{Lisa Altinier}
\author[d]{Ramya Anche}
\author[h]{Pierre Baudoz}
\author[i]{Beth Biller}
\author[f]{Sarah Blunt}
\author[j]{Wolfgang Brandner}
\author[k]{Marah Brinjikji}
\author[h]{\'Oscar Carri\'on-Gonz\'alez}
\author[f]{Amanda Chavez}
\author[g]{Elodie Choquet}
\author[l]{David Doelman}
\author[m]{Julien H. Girard}
\author[n]{Alexandra Z. Greenbaum}
\author[o]{Samantha N. Hasler}
\author[d]{Justin Hom}
\author[n]{James G. Ingalls}
\author[p]{Stephen R. Kane}
\author[q]{N. Jeremy Kasdin}
\author[j]{Oliver Krause}
\author[r,s]{Masayuki Kuzuhara}
\author[g]{Alexis Lau}
\author[p]{Zhexing Li}
\author[r,s]{John Livingston}
\author[n]{Patrick J. Lowrance}
\author[t]{Kevin Ludwick}
\author[u]{Bruce Macintosh}
\author[c]{Eric Mamajek}
\author[v]{Mark Marley}
\author[h]{Johan Mazoyer}
\author[c]{Bertrand Mennesson}
\author[r,s]{Toshiyuki Mizuki}
\author[v]{Sarah E. Moran}
\author[w]{Naoshi Murakami}
\author[s,x,r]{Jun Nishikawa}
\author[f]{Malachi Noel}
\author[y]{Laurent Pueyo}
\author[c]{Sergi Hildebrandt Rafels}
\author[c]{Jason Rhodes}
\author[v]{Tyler Robinson}
\author[z]{Robert J. De Rosa}
\author[j]{Matthias Samland}
\author[d]{Nicholas Schragal}
\author[j]{J\"urgen Schreiber}
\author[n]{Jennifer Sobeck}
\author[c]{Karl Stapelfeldt}
\author[r,s]{Motohide Tamura}
\author[aa]{Taichi Uyajma}
\author[g]{Arthur Vigan}
\author[bb,cc]{Michele Woodland}
\author[c]{Marie Ygouf}
\author[s]{Kenta Yoneta}
\author[cc]{Robert T. Zellem}
\author[cc]{Neil T. Zimmerman}
\affil[c]{Jet Propulsion Laboratory, California Institute of Technology, 4800 Oak Grove Drive, Pasadena, CA 91109, USA}
\affil[d]{Steward Observatory and the Department of Astronomy, The University of Arizona, 933 N Cherry Ave, Tucson, AZ, 85721, USA}
\affil[e]{University of California, Santa Barbara, California, 93106, USA}
\affil[f]{Center for Interdisciplinary Exploration and Research in Astrophysics (CIERA) and Department of Physics and Astronomy, Northwestern University, Evanston, IL 60208, USA}
\affil[g]{Aix Marseille Univ, CNRS, CNES, LAM, Marseille, France}
\affil[h]{LESIA, Observatoire de Paris, Universit\'e PSL, CNRS, Sorbonne Universit\'e, Universit\'e Paris Cit\'e, Meudon, France}
\affil[i]{Institute for Astronomy, University of Edinburgh, EH9 3HJ, Edinburgh UK}
\affil[j]{Max Planck Institute for Astronomy, K\"onigstuhl 17, 69117 Heidelberg, Germany}
\affil[k]{School of Earth and Space Exploration, Arizona State University, 781 E Terrace Mall, Tempe, AZ 85287-6004}
\affil[l]{SRON Netherlands Institute for Space Research, Niels Bohrweg 4, 2333 CA, Leiden, The Netherlands}
\affil[m]{Space Telescope Science Institute, 3700 San Martin Drive, Baltimore, MD 21218, USA}
\affil[n]{IPAC, Caltech, 1200 E. California Blvd., Pasadena, CA 91125}
\affil[o]{Department of Earth, Atmospheric and Planetary Sciences, Massachusetts Institute of Technology, 77 Massachusetts Ave., Cambridge, MA 02139}
\affil[p]{University of California, Riverside}
\affil[q]{Princeton University, Princeton, New Jersey, 08544}
\affil[r]{Astrobiology Center, NINS, 2-21-1 Osawa, Mitaka, Tokyo 181-8588, Japan}
\affil[s]{National Astronomical Observatory of Japan, NINS, 2-21-1 Osawa, Mitaka, Tokyo 181-8588, Japan}
\affil[t]{University of Alabama in Huntsville, 301 Sparkman Dr, Huntsville, AL 35899, USA}
\affil[u]{University of California Santa Cruz, 1156 High Street Santa Cruz CA 95064}
\affil[v]{Lunar \& Planetary Laboratory, University of Arizona, Tucson, AZ 85721 USA}
\affil[w]{Hokkaido University, Kita-13, Nishi-8, Kita-ku, Sapporo, Hokkaido 060-8628, Japan}
\affil[x]{The Graduate Universtiy for Advanced Studies (SOKENDAI), 2-21-1 Osawa, Mitaka, Tokyo 181-8588, Japan}
\affil[y]{STScI, 3700 San Martin Drive, Baltimore, MD 21230}
\affil[z]{European Southern Observatory, Alonso de C\'{o}rdova 3107, Vitacura, Santiago, Chile}
\affil[aa]{Department of Physics and Astronomy, California State University Northridge, 18111 Nordhoff Street, Northridge, CA 91330, USA}
\affil[bb]{Southeastern Universities Research Association, Washington DC, 20005, USA}
\affil[cc]{NASA's Goddard Space Flight Center, 8800 Greenbelt Rd., Greenbelt, MD 20771}

\authorinfo{Send correspondence to Dmitry Savransky 
\texttt{ds264@cornell.edu}}

\begin{document} 
\maketitle

\begin{abstract}
In preparation for the operational phase of the Nancy Grace Roman Space Telescope, NASA has created the Coronagraph Community Participation Program (CPP) to prepare for and execute Coronagraph Instrument technology demonstration observations. The CPP is composed of 7 small, US-based teams, selected competitively via the Nancy Grace Roman Space Telescope Research and Support Participation Opportunity, members of the Roman Project Team, and international partner teams from ESA, JAXA, CNES, and the Max Planck Institute for Astronomy. The primary goals of the CPP are to prepare simulation tools, target databases, and data reduction software for the execution of the Coronagraph Instrument observation phase. Here, we present the current status of the CPP and its working groups, along with plans for future CPP activities up through Roman's launch. We also discuss plans to potentially enable future commissioning of currently-unsupported modes.
\end{abstract}


\section{INTRODUCTION}\label{sec:intro} 
The Nancy Grace Roman Space Telescope (Roman) is NASA's next major Astrophysics Observatory, scheduled for launch in October 2026. Roman is composed of a wide-field, 2.4 m near-infrared-optimized telescope equipped with a precision camera, the Wide-Field Instrument, and
a Coronagraph Instrument technology demonstration.  The Coronagraph Instrument will provide the first-ever, in-space demonstrations of multiple key technologies that will be required by any future mission whose goals include the direct imaging and characterization of rocky planets in the habitable zones of nearby, sun-like stars. These technologies include:
\begin{itemize}
\item Ultra-precise wavefront sensing and control;
\item Large-format deformable mirrors;
\item High-contrast coronagraphs operating with obscured and complex primary geometries; and
\item Ultra-low-noise, photon-counting, EMCCD detectors.
\end{itemize}
In addition to these key technologies, Coronagraph Instrument data will provide us with the first opportunity to apply advanced data post-processing techniques to space imaging data in a relevant contrast regime.

While the Coronagraph Instrument will not itself be sensitive to terrestrial planets, it will be capable of imaging giant planets in reflected, visible light, and will potentially even be sensitive to Jovian analogues for a limited number of sufficiently bright and nearby targets.  Such targets will be selected from the pool of known, self-luminous, young planets discovered via ground-based, near-IR imaging\cite{lacy2020prospects} as well as those discovered via radial velocity measurements\cite{batalha2018color,saxena2021simulating}. There is also the potential that additional imaging targets will be provided via astrometric discoveries from Gaia\cite{prusti2016gaia}.

The Corongraph Instrument will also be capable of imaging low-surface brightness disks at smaller working angles than currently possible with the Hubble Space telescope, allowing us to learn more about their morphology\cite{mennesson2018wfirst}.  At the same time, the Roman Coronagraph has the potential to provide us with the first-ever visible-light images of exozodiacal dust\cite{douglas2022PASP}.  This is particularly important for informing the development of a dedicated, future, exoplanet-imaging mission, such as the Habitable Worlds Observatory (HWO), as the average brightness of exozodiacal light in visible wavelengths is currently unknown, and could potentially be a limiting background signal for Earth-like exoplanet detection. 

 Crucially, the Roman Coronagraph will demonstrate end-to-end observing operations with an instrument designed for exoplanet coronagraphy,  in the presence of actual
observatory disturbances and astrophysical noise sources, making it an indispensable step on the way to HWO.

The Roman Coronagraph Community Participation Program (CPP) was established by the 2022 NASA ROSES Nancy Grace Roman Space Telescope Research and Support Participation Opportunities program element. This call solicited proposals aimed at supporting the progress of and exploiting the scientific and technical data from Roman.  Three categories of proposals were solicited: Wide Field Science, Project Infrastructure Teams, and Coronagraph Community Participation.  The CPP category was limited to sole-PI or small-group proposals, with the intent to form a single, cohesive CPP Team from the selected teams, as well as international partners and representatives from the Coronagraph Instrument science and engineering teams from NASA's Jet Propulsion Laboratory and Goddard Space Flight Center, and members of the Roman Science Support Center (SSC) at Caltech/IPAC. The SSC is part of the Roman Ground System and is responsible for  supporting Coronagraph Instrument operations in addition to the implementation of science data pipelines for the Wide Field Instrument spectroscopy and the Galactic Bulge Time Domain Survey. 

\section{Roman Coronagraph Overview}
Extensive, detailed descriptions of the Roman Observatory and the Coronagraph Instrument are available in the literature, and especially in Refs.~\citenum{spergel2015wide,akeson2019wide, bailey2023SPIE}.  A definitive description of the coronagraph design is available in \refnum{riggs2021flight}, and details on its optical modeling can be found in \refnum{krist2023end}. An extensive review of past work on simulation and performance modeling for the Coronagraph Instrument can be found in \refnum{douglas2020review}.  Finally, the results of the instrument-level test program for the Coronagraph Instrument are available in \refnum{poberezhskiy2024roman}.  Here, we will focus on describing those aspects of Roman and the Coronagraph Instrument and its operations concept that are of particular importance to CPP activities. 

\subsection{Roman Field of Regard}

For CGI observations, the Roman observatory must be oriented such that the angle between the boresight (telescope to target star vector) and the telescope-sun vector must be between 56$^\circ$ and 124$^\circ$. The placement of the solar panels (parallel to the boresight) places an additional roll constraint of $\pm 13^\circ$ about the boresight to ensure proper illumination of the panels (a $0^\circ$ roll is defined as orienting the solar panels such that they are orthogonal to the telescope-sun vector). Additionally there is a 22$^\circ$ Earth avoidance and 11$^\circ$ moon avoidance pointing constraint. Jointly, these constraints mean that targets within $\sim 34^\circ$ of the ecliptic poles are continuously observable, whereas targets near the ecliptic can only be observed for approximately half the year (and in some cases, as little as 150 days of each year, depending on the specific observatory orbit geometry). These values may be relaxed by 1-2$^\circ$ with an update of the final power margins, however, the average target availability will remain qualitatively the same.

\begin{figure}[ht]
\centering
\includegraphics[width=\textwidth]{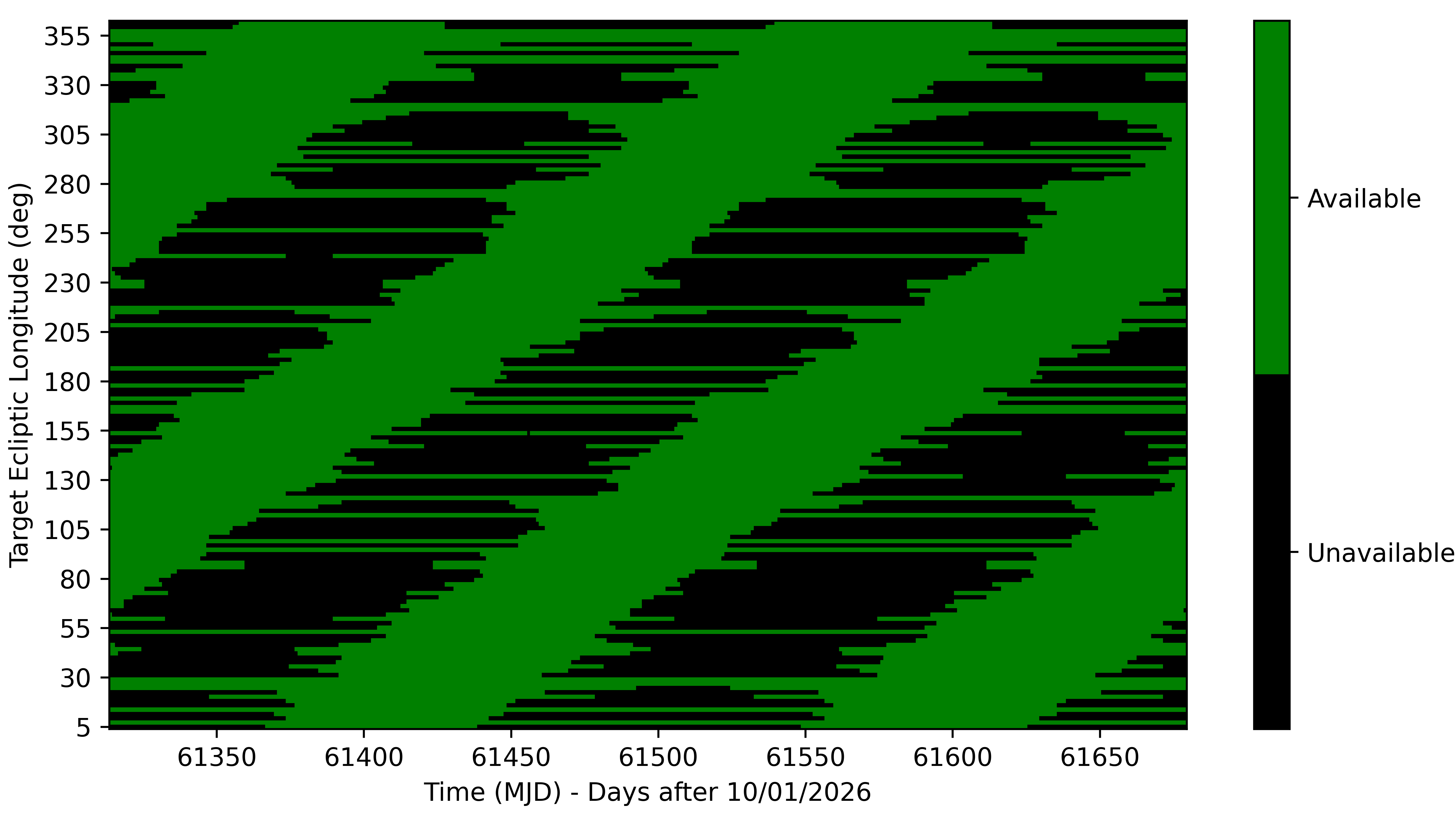}
\caption{Roman observatory pointing constraints applied to the HWO mission star list\cite{mamajek2024nasa}.  Rows represent individual target stars (sorted by ecliptic longitude) while columns represent time in days after 10/01/2026. Black regions of each row represent times when the target is inaccessible due to pointing constraints. The fraction of the year that a target is available is primarily driven by its ecliptic latitude.}\label{fig:cgi_keepout}
\end{figure}

\reffig{fig:cgi_keepout} shows a visualization of Roman's pointing constraints on target availability for the 164 stars in the Habitable Worlds Observatory mission star list\cite{mamajek2024nasa}.  It is important to note that this list was selected for visualization purposes only, and is not the target list that will be used for Roman Coronagraph observations. However, the HWO mission star list is primarily composed of bright, nearby stars, making it representative of the expected spread of Roman Coronagraph targets on the sky. The visualization in \reffig{fig:cgi_keepout}, sometimes referred to as a keepout map, shows the availability of each target as a function of time.  Targets are ordered by their ecliptic longitude, which produces the systematic pattern of availability for those targets that are not observable year-round.  The fraction of the year that targets are available for observation is driven primarily by the targets' ecliptic latitudes, as detailed above, and secondarily by the specifics of the observatory orbit.  The final schedule for the Coronagraph observation phase will need to be fully consistent with the final target lists' observing availability. 

\subsection{Roman Orbit}

The Roman observatory will be placed on a halo-like orbit about the Sun-Earth L$_2$ point.  The orbit will be periodic in the sun-Earth rotating frame, dipping above and below the ecliptic approximately every 3 months. However, the specific final observatory orbit will be driven entirely by the launch date. 
Roman can launch starting in October 2026 (when the launch vehicle is to be delivered) and is expected to launch no later than March of 2027.  While the orbits that are accessible from an October vs. March launch belong to the same family of periodic L$_2$ structures, they are qualitatively different, with different magnitudes of their out-of-ecliptic components to the north and south.  Furthermore, the phasing onto these orbits will similarly depend on the launch date. These differences in mission orbits induce differences in the keepout maps, meaning that the Coronagraph instrument's observing schedule can only be finalized once the launch date is known exactly (and will likely need to be re-evaluated during the mission's cruise phase on the way to the final orbit). 


\subsection{Coronagraph Observing Sequence}

A Roman Coronagraph high-contrast observing sequence will include observations of two stars: the target star and a bright reference star.  The full observing sequence will consist of observations of the reference star, which will be used both to collect reference images for post-processing as well as to create the regions of high-contrast via iterative wavefront control, followed by observations of the target star at multiple telescope roll angles, followed by a return to the reference star.  The reference star requirements are that they be bright (V band magnitudes of less than 3), have small angular diameters ($< 2$ mas), must be single stars without known planets or other bound objects of their own, and must have a small solar pitch difference with the target star (the specific final requirement will be established with future analysis to be completed by the Coronagraph instrument team and CPP). A full description of the observing sequence and additional discussion on reference star criteria is available in \refnum{krist2023end}.  Collectively, these criteria rule out the vast majority of all bright stars, leaving fewer than 100 potential references.  Further vetting of these will likely reveal that some are multiples, meaning that the availability of reference stars will be one of the most challenging constraints to scheduling the Coronagraph observing program. This makes the construction of the reference star catalog one of the highest priority goals for the CPP (see \refsec{sec:obswg}). 

\subsection{Ground-in-the-Loop Wavefront Control}

The operations concept for the Coronagraph Instrument's wavefront control scheme is that data processing will be done on the ground, known as Ground-in-the-Loop (GITL).  Reference star observations used for wavefront control will be downlinked to Earth, transmitted to the SSC, where they will be analyzed and used to generate new wavefront control commands to the deformable mirrors, which will then be uplinked back to the telescope.  This imposes multiple constraints on Coronagraph observations.  Most importantly, these will need to be fully planned weeks to months in advance to ensure ground station contact during high-order wavefront sensing (HOWFS) activities.  Similarly, there will be no opportunity for mid-observation changes for Coronagraph observations.  On the other hand, the GITL infrastructure leaves open the possibility for new, contributed HOWFS modes once the observatory is on sky.  If data processing were to be done on the spacecraft, new modes could only be enabled via observatory software updates, which can be challenging to implement and execute. However, any algorithm that can be sufficiently demonstrated on the ground (with specific demonstration criteria to be determined in the future by the project) and can be made consistent with the GITL architecture has the potential for future use with the Roman Coronagraph.

\subsection{Coronagraph Observing Modes}

As the Coronagraph Instrument is a class D technology demonstration, it will launch with hardware components that have not been fully characterized and/or tested.  This means that the Coronagraph has three classes of observing modes: required (the Band 1 narrow field-of-view imaging mode with the hybrid Lyot coronagraph), best-effort (wide field-of-view imaging and spectroscopy with shaped pupil coronagraphs and polarimetric imaging), and unsupported (all other modes).  The required mode has been fully tested and commissioned\cite{poberezhskiy2024roman}.  All other modes have varying levels of ground-testing and support.  This again leaves open the possibility that multiple modes, including formally unsupported ones, can be commissioned on-sky.  A detailed description of all Coronagraph Instrument masks and modes is available in \refnum{riggs2021flight}.

\section{CPP Structure and Working Groups}

The CPP is chaired by the Roman Coronagraph Instrument Technologist. A co-chair is appointed by the chair for a term of roughly one year. The CPP Core Team consists of the 7 US team PIs selected in the original ROSES program call, the 4 PIs named by the international partners, and the Roman project ex officio members, which includes the JPL project scientist, the JPL deputy project scientists, the Goddard coronagraph scientist, and the SSC scientist. The Core Team is charged with advising the chair and co-chair in establishing and enforcing team structure and policies, and setting team goals and priorities. Other members of the US and international partner teams form the CPP Project Team, which carries out the work of the CPP.  Beyond this, the CPP Community Team includes anyone wishing to work with the CPP, but not committed to supporting high-priority CPP activities. Finally, any member of the interested community is welcome to attend public CPP and Roman events.  The CPP is governed by a collaboration agreement\footnote{\nlhref{https://docs.google.com/document/d/1Z2FAGD3KSqiNx5jbS_iIG1dv8aZZPJ8ENeo9EMBFH0s}}.  The collaboration agreement establishes three initial working groups for carrying out the activities of the CPP, while also allowing for additional working groups to be instantiated to meet evolving project goals and priorities (see \refsec{sec:next}).

\subsection{Observation Planning Working Group}\label{sec:obswg}
The activities of the observation planning working group are detailed in \refnum{wolff2024roman}. This group’s activities encompass target selection, target database development and maintenance, precursor observations, reference star identification, development and maintenance of the exposure time calculator, astrophysical modeling, and observation planning for both primary and unsupported modes.  The observation planning working group's initial activities have been primarily focused on the establishment of the reference star catalog and the accompanying vetting program of all potential reference stars.

\subsection{Data Reduction and Simulation Working Group}
The activities of the data reduction and simulation working group are detailed in \refnum{millar2024roman}.  This group’s activities will include the development of the coronagraph instrument required mode data reduction pipeline (DRP), generating image simulations and performance predictions, and developing algorithms for data analysis and calibration. DRP development is the most important initial focus of this working groups, with an initial release planned for September 2024 (see \refnum{millar2024roman} for details and release schedule).  The DRP is being developed as an open-source project, accepting contributions from both the CPP and the broader community\footnote{see: \nlhref{https://github.com/roman-corgi/corgidrp}}.

\subsection{Hardware Working Group}
This group’s activities include assisting in integration, testing, commissioning, and operations of any and all observing modes to be used on sky, assisting with data analysis of integration and test data, and researching high-order wavefront sensing and control algorithms. 

\subsection{Climate Committee}
In addition to the working groups, the CPP collaboration agreement establishes a climate committee, which is focused on achieving inclusivity within the CPP, as guided by NASA's core values. To that end, the Climate Committee has developed a CPP-wide Inclusion Plan and maintains a suite of inclusivity-related resources for CPP teams and members. The Inclusion Plan includes an emphasis on empowering Early Career Researchers within CPP and also commits the Climate Committee to engaging in an annual CPP-wide climate survey.

\section{CPP Membership}\label{sec:membership}

The current CPP membership consists of the US-based teams selected via the initial ROSES program element, along with representatives from NASA centers, international partner teams, the SSC, and volunteers from the broader community. 

\subsection{US Teams}
Seven US institution-based teams were selected in the first call of the NASA ROSES Nancy Grace Roman Space Telescope Research and Support Participation Opportunities in the CPP category\footnote{The full abstracts of the selected investigations are available at \nlhref{https://nspires.nasaprs.com/external/viewrepositorydocument/cmdocumentid=950761/solicitationId=\%7B1BD0AA55-40BB-1419-EEA1-64FF5B4269D3\%7D/viewSolicitationDocument=1/Roman-22_Abstracts_20230706_Benford.pdf}}. The US teams are based at institutions of higher learning and NASA centers spread throughout the United States.  The teams have a broad range of relevant expertise and focus areas, including coronagraph hardware, advanced high-order wavefront sensing and control, multi-star wavefront control, data reduction and post-processing, instrument calibration, atmospheric modeling, and observation scheduling.  Several of the US teams are also involved with planning for HWO, and plan to investigate how lessons learned from the Roman Coronagraph apply to HWO.

\subsection{International Partners}
International participation in the CPP is provided by partner teams from ESA, JAXA, CNES, and the Max Planck Institute for Astronomy (MPIA). The MPIA team has active participants in all CPP working groups  and also has plans for further engagement with the German science community and the Council of German Observatories (Rat deutscher Sternwarten; RDS). ESA is represented by two appointed scientists who are tasked with representing the interests of the European scientific community in the Roman mission. JAXA has already assembled a large team to support CPP activities, spread through all working groups. JAXA will help lead many of the CPP polarimetry efforts, and will assist in the search for suitable Coronagraph targets via observing programs at the Subaru observatory. CNES are engaging in all three working groups and the leads are tasked with engaging the broader French high-contrast community.
 
\section{STATUS AND NEXT STEPS}\label{sec:next}
The CPP has been operating since Fall 2023 and currently includes over 90 members in the core, project, and community teams.  To date, the CPP has established three active working groups, which have spearheaded a number of key activities, including the development of the Coronagraph Instrument data reduction pipeline, the submission of numerous observing proposals for reference star vetting (several of which have been accepted), and the initial establishment of observing plans for the Coronagraph's observation phase. At the same time, the CPP continues to grow and evolve to meet the overall project's needs.  A new polarimetry working group, co-led by JAXA CPP members will be established to coordinate polarimetry-related CPP activities. Similarly, the CPP continues to engage with the broader community, as well as with researchers working towards HWO and other future missions to help ensure that the Roman Coronagraph will pave the way for a dedicated, exoplanet imaging mission.

\acknowledgments 
This work was supported by NASA under award Nos. 80NSSC24K0216, 80NSSC24K0217, 80NSSC24K0087, 80NSSC24K0097, 80NSSC24K0085, and 80NSSC24K0218. The CPP gratefully acknowledges the contributions of the $>1600$ scientists and engineers that have helped make the Coronagraph Instrument possible. 

\bibliography{local}   
\bibliographystyle{spiebib}

\end{document}